\documentclass[preprint,10pt]{elsarticle}
\usepackage{svg}

\usepackage{amssymb}
\usepackage{amsmath}
\usepackage{threeparttable}
\usepackage[english]{babel}
\usepackage[T1]{fontenc}
\usepackage[utf8]{inputenc}
\usepackage{multirow}
\usepackage{array}
\usepackage{threeparttable}
\usepackage{booktabs} 
\usepackage[table]{xcolor}      

\usepackage{acronym}

\usepackage{graphicx}
\usepackage{subcaption} 

\usepackage{hyperref}
\hypersetup{
  colorlinks   = true, 
  urlcolor     = blue, 
  linkcolor    = black, 
  citecolor   = blue 
}

\usepackage{datatool}

\usepackage{setspace}

\journal{Elsevier}

\begin{document}

\begin{frontmatter}



\title{\textbf{
Real-Time Physics-Aware Battery Health Monitoring from Partial Charging Profiles via Physics-Informed Neural Networks
}}

\author{Xubo Gu$^{a}$, Xun Huan$^{b}$, Yao Ren$^{c}$, Wenqing Zhou$^{c}$, Weiran Jiang$^{c*}$, and Ziyou Song$^{a,d*}$}

\cortext[mycorrespondingauthor]{Corresponding Authors:  ziyou@umich.edu (Z. Song), wjiang@farasis.com (W. Jiang)}

\address{$^a$Department of Mechanical Engineering, National University of Singapore, Singapore, 117575, Singapore}
\address{$^b$Department of Mechanical Engineering, University of Michigan, Ann Arbor, MI, 48109, USA}
\address{$^c$Farasis Energy USA, Inc., Hayward, CA, 94545, USA}
\address{$^d$Department of Electrical Engineering and Computer Science, University of Michigan, Ann Arbor, MI, 48109, USA}

\begin{abstract}\small
Monitoring battery health is essential for ensuring safe and efficient operation. However, there is an inherent trade-off between assessment speed and diagnostic depth---specifically, between rapid overall health estimation and precise identification of internal degradation states.
Capturing detailed internal battery information efficiently remains a major challenge, yet such insights are key to understanding the various degradation mechanisms. 
To address this, we develop a parameterized physics-informed neural network (P-PINNSPM) over the key aging-related parameter space for a single particle model.
The model can accurately predict internal battery variables across the parameter space and identifies internal parameters in about 30 seconds---achieving a $47 \times$ speedup over the finite volume method---while maintaining high accuracy.
These parameters improve the battery state-of-health (SOH) estimation accuracy by at least 60.61\%, compared to models without parameter incorporation.
Moreover, they enable extrapolation to unseen SOH levels and support robust estimation across diverse charging profiles and operating conditions.
Our results demonstrate the strong potential of physics-informed machine learning to advance real-time, data-efficient, and physics-aware battery management systems.
\end{abstract}




\end{frontmatter}
\setstretch{1.5}
\section{Introduction}

Li-ion batteries are critical enablers for electric transportation and renewable energy storage due to their high energy density and long lifespan. However, electrochemical reactions progressively degrade battery performance, raising safety concerns and operational costs over time \cite{gu_challenges_2024}. These challenges highlight the importance of battery health diagnostics. Accurate health monitoring not only ensures safety and reliability but also enables performance optimization and cost reduction.

Battery health diagnostics are handled by the battery management system (BMS), which monitors key parameters to maintain safe and efficient operation. Empirical and equivalent circuit models remain the dominant algorithms used in BMSs \cite{lyu2020li} due to their computational efficiency. However, these methods offer only cell-level information, such as the state of charge (SOC) and state of health (SOH), and do not provide insights into the internal condition of the battery. This can be a limitation, as two cells with similar SOH values may exhibit vastly different internal states and degradation rates.

Data-driven methods have gained significant traction for SOH estimation by learning patterns directly from historical data, bypassing complex governing equations \cite{ZHOU2024100346}. 
These approaches typically extract features from voltage, current, and temperature profiles, and offer fast execution \cite{LU2024100338, LU2024100361, MAO2025236982}. 
However, they often operate as opaque ``black-boxes'' models, lacking interpretability and physical insight. 
Mechanical signals such as pressure and expansion can reflect internal battery states and offer physical insights into aging \cite{gu2025mechanical, DING2025103998}. However, obtaining such information requires implanting microsensors within the cells in real-life applications.

Physics-based models, in contrast, provide a structured framework for tracking internal battery states. First-principles models such as the Doyle--Fuller--Newman (DFN) model describe transport and reaction processes in Li-ion batteries through partial differential equations (PDEs).
A simplified variant, the single particle model (SPM), represents each electrode with a single spherical porous particle and neglects the electrolyte dynamics.
Despite their interpretability and predictive power, these models are computationally intensive, limiting their practical use in real-time diagnostics.

Hybrid models that combine physics-based modeling with data-driven approaches have emerged to balance accuracy and efficiency \cite{KIM2023100293, LIU2025100446, tian2024exploiting, tu_integrating_2023}. For example, Kohtz et al. \cite{kohtz_physics-informed_2022} used a finite element model to infer SEI thickness from voltage curves, combining it with other electrical features for SOH estimation using Gaussian process regression.
Elsewhere, Hofmann et al. \cite{hofmann_physics-informed_2023} improved SOH prediction by incorporating internal states derived from the DFN model.
While effective, these methods remain computationally demanding, creating a bottleneck for real-time implementation in BMSs.
This highlights a critical challenge: {how to retain the interpretability of physics-based models while enabling fast evaluation of internal states.}

Physics-informed neural networks (PINNs) offer a promising solution. By embedding physical laws---such as governing PDEs and boundary conditions---into the neural network loss function and optimizing via automatic differentiation, PINNs allow efficient solution of complex physical models \cite{karniadakis_physics-_2021, raissi_physics-informed_2019}. 
Recent studies have explored PINNs for battery modeling, with varying levels of physical fidelity and practicality.
For example, Wang et al. proposed a PINN framework for SOH prediction across different battery types and conditions; however, the lack of explicit physics limited its interpretability \cite{wang_physics-informed_2024-2}.
Zheng et al. used physics-informed deep operator networks to map current curves to terminal voltage, incorporating solid diffusivity and mechanical stress \cite{zheng_inferring_2023}, and later extended this to a physics-informed multiple-input operator network for state-space modeling of an enhanced SPM \cite{zheng_state-space_2023}. 
Huang et al. developed a model-integrated neural network for DFN modeling that demonstrated good extrapolation to unseen profiles \cite{huang_minn_2024}. 
While these approaches improve efficiency, they often fail to account for aging-related parameters. This omission reduces accuracy as the cell degrades, making them unsuitable for long-term diagnostics. 
Models intended for real-world applications must incorporate aging-related parameters to remain valid across the battery's lifetime.

Recent works address the gap in incorporating parameters into battery models. Hassanaly et al. incorporated parameter identification into PINNs for both the SPM and DFN models, though their parameter selection was limited and primarily demonstrative \cite{hassanaly_pinn_2024-1, hassanaly_pinn_2024}. 
Mendez et al. encoded two SPM parameters into a PINN and showed promising prediction performance, but did not explore downstream applications \cite{mendez-corbacho_physics-informed_2024}.
Notably, both studies focused on constant-current discharging profiles---scenarios that are unrealistic in real-world applications. In contrast, charging profiles offer more controllability and are widely available in real-world settings, making them a better basis for health diagnostics. Besides, both studies failed to select the most influential parameters as cell ages.

Taken together, these limitations point to a key research gap: the need for a PINN-based hybrid model that can (1) embed aging-related parameters, (2) enable fast and accurate estimation of internal battery states from charging profiles, and (3) support improved SOH diagnostics over the battery's lifetime.

The novelty and contributions of this work are as follows:
\begin{itemize}
\item We present a systematic approach for selecting key aging-related parameters in SPM by identifying independent variables and sensitivity analysis. 
\item We develop a parameterized PINN for the SPM that we call P-PINNSPM,  incorporating the identified key parameters to efficiently and accurately infer internal battery states across the parameter space.
\item We conduct comprehensive experiments demonstrating SOH estimation accuracy. 
The model generalizes well to previously unseen SOH levels, varying current profiles, and different working conditions.
\end{itemize}

The remainder of this paper is organized as follows.
Section \ref{sec:method} describes the methods and datasets used in this study.
Section \ref{sec:res} presents the experimental results, including improved SOH estimation enabled by internal parameters and the extrapolation capabilities of P-PINNSPM.
Section \ref{sec:con} concludes the paper and outlines directions for future research.

\section{Methods and datasets}\label{sec:method}

\begin{figure*}
    \centering
    \includegraphics[width=0.85\linewidth]{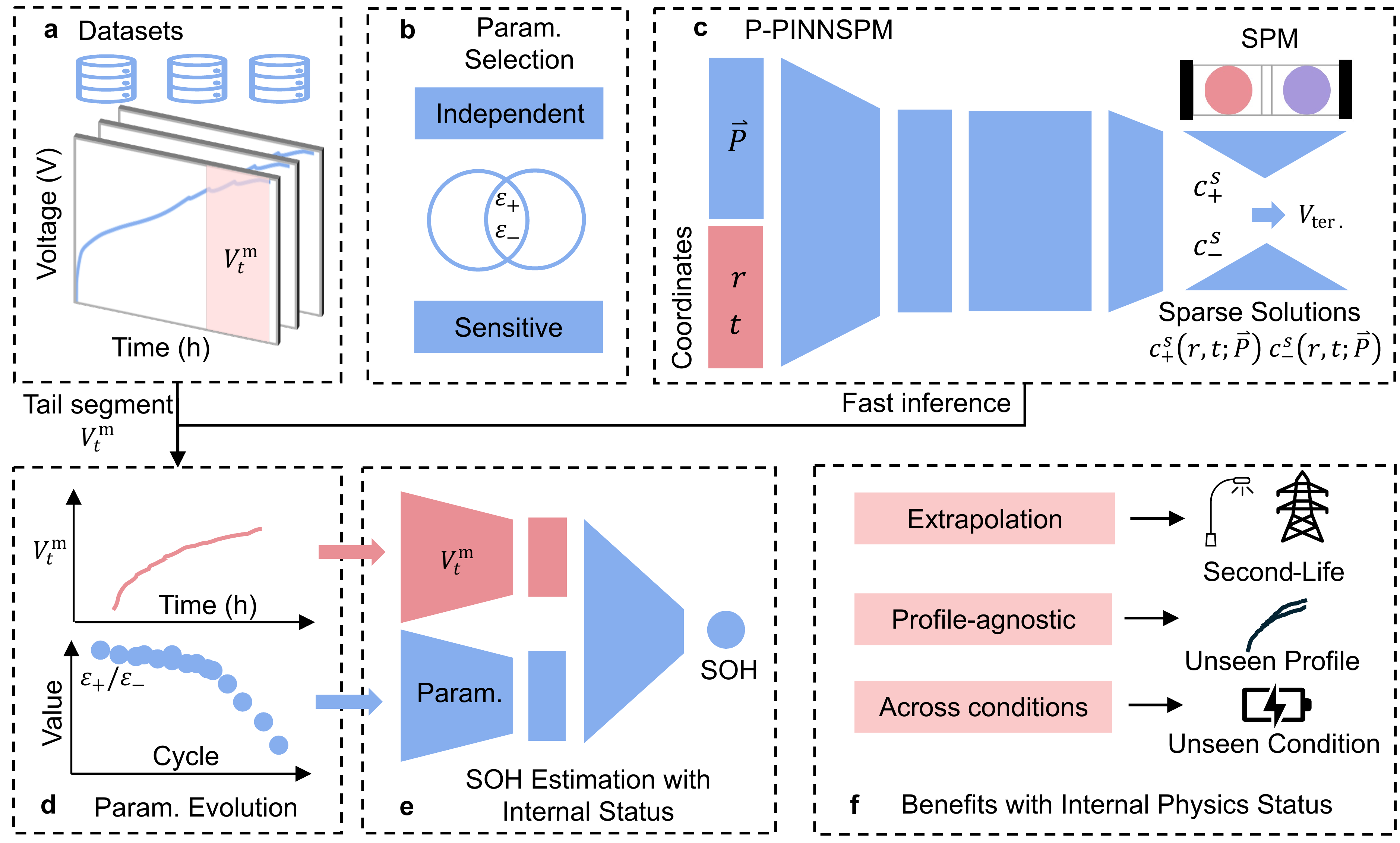}
    \caption{The technical framework for real-time battery health diagnostic with internal physical status.}
    \label{fig:diagram}
\end{figure*}

Figure \ref{fig:diagram} provides an overview of the study.
We begin by testing and collecting data from large-format commercial pouch cells under four distinct operating conditions, as shown in Table \ref{tab:2}.
In this study, we use only the tail charging voltage segment (Figure \ref{fig:diagram}a) to estimate SOH, since tail-end profiles are more stable and widely available in practice. This choice improves the applicability of the method to real-world scenarios.

A key challenge is deciding which parameters from the SPM to include in the PINN. We identify the minimal set of independent parameters required to fully characterize the model and evaluate their sensitivity to the voltage response (Figure \ref{fig:diagram}b). This analysis guides the selection of key aging-related parameters.

These selected parameters are then embedded into a parameterized PINN (P-PINNSPM as shown in Figure \ref{fig:diagram}c). The network spans the whole parameter space alongside spatial and temporal coordinates. By integrating battery physics, it can infer internal states across a wide range of parameter values.

To track parameter evolution over time, we employ a differential evolution algorithm. With the fast inference capability of P-PINNSPM, we can obtain real-time updates of key internal parameters (Figure \ref{fig:diagram}d). These identified parameters are then used as inputs to a dedicated SOH estimation network, which includes separate encoders for time-series voltage data and internal parameters (Figure \ref{fig:diagram}e).

The results show that incorporating physical parameters not only improves SOH estimation but also enables robust prediction of unseen SOH levels. The approach further supports profile-agnostic estimation within a cell and generalizes well across different cells and operating conditions (Figure \ref{fig:diagram}f). This highlights the strong potential of physics-informed machine learning to advance real-time, data-efficient, and physics-aware battery management systems.

The following subsections provide a step-by-step explanation of each component in the proposed framework.

\subsection{Datasets}
\begin{figure}[!t]
    \centering
    \begin{subfigure}[b]{0.47\textwidth}
        \centering
        \includegraphics[width=\linewidth]{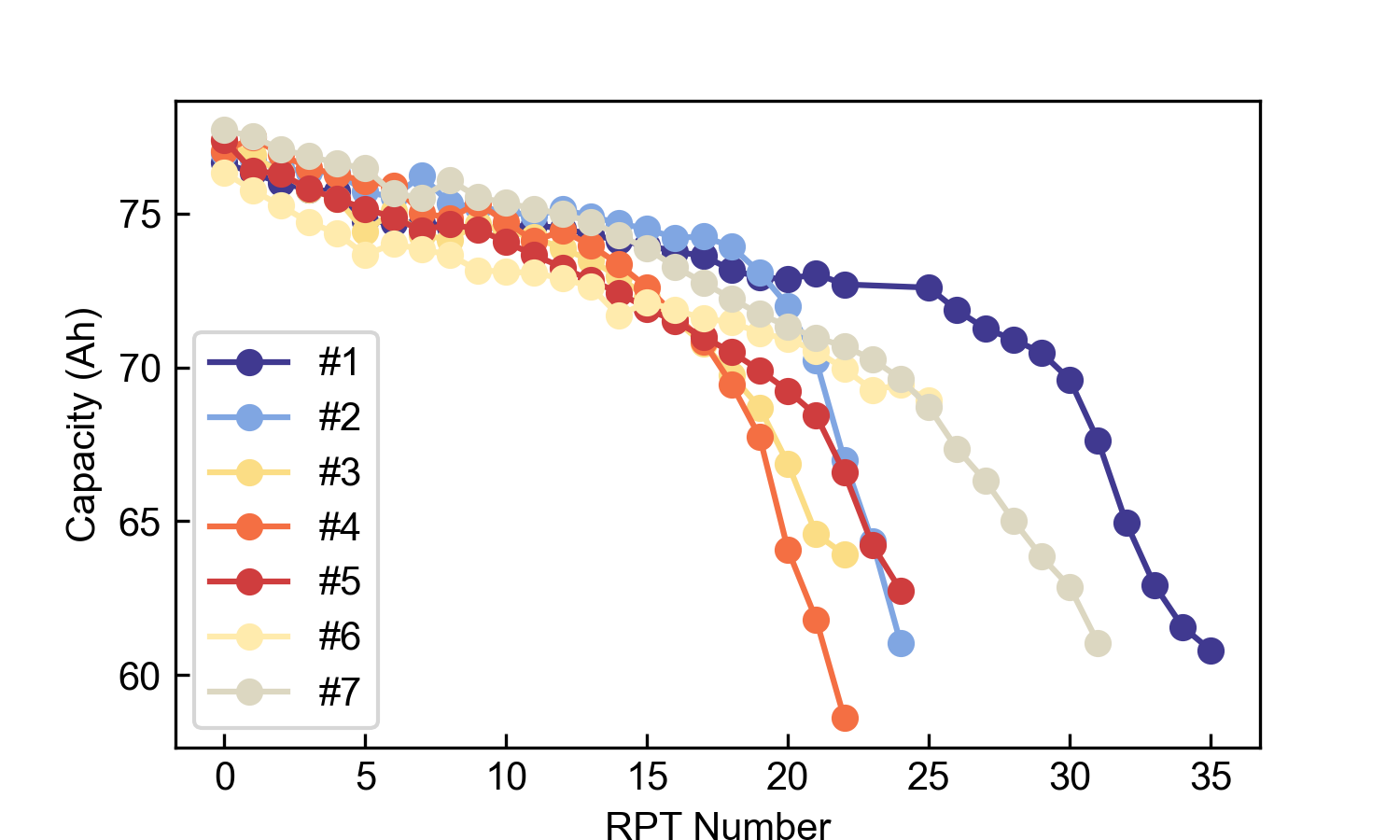}
        \caption{}
        \label{fig:cap_sum}
    \end{subfigure}
    \hfill
    \begin{subfigure}[b]{0.47\textwidth}
        \centering
        \includegraphics[width=\linewidth]{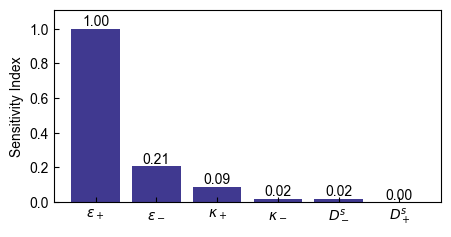}
        \caption{}
        \label{fig:SA_res}
    \end{subfigure}
    \caption{Battery capacity fade and parameter sensitivity. (a): Battery capacity fade with respect to RPT cycle number; (b): Sensitivity analysis for independent parameters.}
    \label{fig:combined}
\end{figure}

We tested and collected data from large-format commercial pouch cells. Seven cells with NCM811/Graphite chemistry and a nominal capacity of 76 Ah were tested under 25$^\circ$C and four types of current profiles, as listed in Table \ref{tab:2}. Every 70 cycles, a reference performance test (RPT) using a C/3 CCCV profile was conducted to calibrate the capacity. Figure \ref{fig:cap_sum} shows the capacity degradation over cycles for all cells.

\begin{table}[h!]
  \centering
  \small
    \caption{Aging test information.}
  \begin{tabular}{
    >{\centering\arraybackslash}m{0.3cm}
    >{\centering\arraybackslash}m{3.5cm}
    >{\centering\arraybackslash}m{3cm}
    >{\centering\arraybackslash}m{1.5cm}
    >{\centering\arraybackslash}m{1.5cm}
  }
    \hline 
    \textbf{No.} & \textbf{C-rate (charge,~discharge)} & \textbf{Operation window (SOC)} & \textbf{Cycle number} & \textbf{Capacity loss (\%)}\\
    \hline 
    \#1 & 0.5 C, 1 C & 0-80\%  & 2558 & 20.68 \\ 
    \#2 & 0.5 C, 1 C & 0-80\%  & 1755 & 20.71  \\
    \#3 & 0.5 C, 1 C & 0-100\% & 1609 & 17.36  \\
    \#4 & 0.5 C, 1 C & 0-100\% & 1609 & 23.92 \\
    \#5 & 1 C, 1 C   & 0-100\% & 1755 & 18.91 \\
    \#6 & 1 C, 1 C   & 0-100\% & 1828 & 9.69 \\
    \#7 & 1 C, 1 C   & 0-80\%  & 2266 & 21.45 \\
    \hline 
  \end{tabular}
  \label{tab:2}
\end{table}

\subsection{Methods}

\subsubsection{Battery model}

\begin{table*}[!t]
\footnotesize
\centering
\caption{Summary of governing equations in the single particle model. 
$i \in {[-,+]}$ denotes the negative or positive electrode.
}
\label{tab:spm_summary}
\renewcommand{\arraystretch}{1.3}
\begin{tabular}{l|l|p{6.5cm}}
\hline
\textbf{Component} & \textbf{Equation} & \textbf{Variable Description} \\
\hline
Solid diffusion &
$\displaystyle \frac{\partial c_i}{\partial t} = \frac{D^s_i}{r^2_i} \frac{\partial}{\partial r_i} \left( r^2_i \frac{\partial c_i}{\partial r_i} \right)$ &
$c_i$: solid-phase Li concentration; $D^s_i$: diffusion coefficient; $r_i$: radial coordinate \\
Boundary conditions &
$\displaystyle \frac{\partial c_i}{\partial r_i}\big|_{r_i = 0} = 0$, 
$\displaystyle D^s_i \frac{\partial c_i}{\partial r_i}\big|_{r_i = R_i} = -j_i$ &
$j_i$: molar flux at particle surface; $R_i$: particle radius \\
Initial condition &
$\displaystyle c_i(0, r_i) = c_i^0$ &
$c_i^0$: initial Li-ion concentration in the particle \\
Flux–current relation &
$\displaystyle j_i = \frac{I}{a_i \delta_i F A}$ &
$I$: applied current; $a_i$: specific surface area; $\delta_i$: electrode thickness; $F$: Faraday constant; $A$: electrode area \\
\hline
Terminal voltage &
$\displaystyle V = U_+\left(\frac{c_+^\text{surf}}{c_+^\text{max}}\right) - U_-\left(\frac{c_-^\text{surf}}{c_-^\text{max}}\right) + \eta_+ - \eta_-$ &
$U_i$: open-circuit potential; $c_i^\text{surf}$: surface concentration; $c_i^\text{max}$: maximum Li concentration in particles; $\eta_i$: overpotential \\
Butler–Volmer kinetics &
$\displaystyle j_i = \frac{2 i_{0,i}}{F} \sinh\left( \frac{F \eta_i}{2RT} \right)$ &
$i_{0,i}$: exchange current density; $\eta_i$: overpotential; $R$: gas constant; $T$: temperature; $F$: Faraday constant \\
\hline
\end{tabular}
\end{table*}

In this work, the SPM is integrated into the PINN. 
Table \ref{tab:spm_summary} summarizes the governing equations of the SPM, and Table~\ref{tab:params} lists the corresponding model parameters.
As a cell ages, identifying and monitoring key parameters becomes crucial.
However, the SPM includes dozens of parameters, but only a few are meaningful for identification. 
This raises the question: which parameters should we prioritize?

To address this, we perform both parameter identifiability and sensitivity analyses. 
The parameter identifiability analysis determines the minimal set of uncoupled parameters that fully characterize the model. This is important since coupled parameters cannot be uniquely inferred from data. 
As verified by Bizeray et al.~\cite{bizeray_identifiability_2019}, six independent parameters can fully represent the SPM: positive/negative particle diffusivity coefficients (\(D^s_+ , D^s_-\)), positive/negative electrode active material volume fraction (\(\varepsilon_+, \varepsilon_- \)), and positive/negative particle reaction rate (\(k_+, k_- \)). 

However, even though all six parameters are theoretically identifiable, their practical identifiability depends on their influence on measurable outputs such as terminal voltage. Some parameters exert only minor effects and therefore more difficult to identify in practice. 
To quantify this, we perform a one-at-a-time sensitivity analysis~\cite{li_parameter_2020, li_data-driven_2022}. Each parameter is varied independently over a \(\pm10\%\) range around its nominal value (as listed in Table~\ref{tab:params}), and 10 values are sampled uniformly. For each variation, we compute the resulting voltage profile, yielding 10 voltage responses per parameter. 

We define the sensitivity index (SI) based on the standard deviation of these responses:
\begin{equation}
    \mathrm{SI} = \sqrt{\frac{1}{N_s} \sum_{j=1}^{N_s}(x_{i,j} - \bar{x}_i)},
    \label{eq:SI}
\end{equation}
where \(\bar{x}_i\) is the average  voltage across the 10 samples for parameter \(i\), and \(N_s=10\) is the number of samples.

The computed SIs are shown in Figure~\ref{fig:SA_res}. The sensitivity ranking, from highest to lowest, is: \(\varepsilon_+ > \varepsilon_- > k_+ > k_- > D^s_- > D^s_+ \). Among all parameters, the positive/negative electrode active material volume fraction exhibits the highest sensitivity to voltage signals. Consequently, we select these two parameters for inclusion in the PINN model.

\subsubsection{PINN}
The application of deep learning to computational science and engineering, especially for solving PDEs, has gained increasing attention in both academia and industry. In settings where acquiring training data is costly or practical, physics-informed machine learning offers a compelling alternative by embedding governing physical laws directly into the learning process \cite{DeRyck_Mishra_2024}. Among such approaches, PINNs have emerged as a prominent framework. 

The key idea behind PINNs is straightforward: since neural networks are universal function approximators, they can be trained to approximate the solution of a PDE by minimizing residuals derived from the underlying physical equations, rather than relying solely on labeled data~\cite{dissanayake1994neural, raissi_physics-informed_2019}. 

Consider the general form of a PDE: 
\begin{equation}
u_t + \mathcal{N}[u] = 0, \quad x \in \Omega, \quad t \in [0, T],
\label{eq:pinn1}
\end{equation}
where \( u(t, x) \) is the unknown solution, \( \mathcal{N}[ \cdot ] \) is a nonlinear differential operator, and \( \Omega \subset \mathbb{R}^D \) is the spatial domain. 
In the PINN framework, the solution \( u(t, x) \) is approximated by a neural network, and derivative terms can be obtained through automatic differentiation \cite{baydin2018automatic}. We define the residual function:
\begin{equation}
f := u_t + \mathcal{N}[u],
\label{eq:pinn2}
\end{equation}
which also becomes a function of the network parameters. The goal is to minimize a composite loss function consisting of three terms:
\begin{equation}
\mathrm{MSE} = \mathrm{MSE}_u + \mathrm{MSE}_f+\mathrm{MSE}_d,
\label{eq:pinn3}
\end{equation}

where \( \mathrm{MSE}_u \) enforces the initial and boundary conditions, \( \mathrm{MSE}_f \) penalizes violations of the governing PDE at a finite set of collocation points in space and time, and \( \mathrm{MSE}_d \) captures discrepancy between network outputs and any available reference data.

\subsubsection{P-PINNSPM and SOH estimation network}\label{sec:Network}
To fully leverage the predictive capabilities of the proposed models, proper training techniques are crucial---especially for the P-PINNSPM, which requires careful architecture design and optimization to achieve strong generalization. To facilitate reproducibility, we provide detailed descriptions of the model structures and training settings for both the P-PINNSPM and the SOH estimation network.
A summary is provided in Table~\ref{tab:net}.

The P-PINNSPM takes as input the active material volume fraction (\(\varepsilon_{-/+}\)), along with spatial and temporal coordinates \((r, t)\). Based on empirical observations, we found that modeling \(c_+\) and \(c_-\) using separate networks improves generalization across the parameter space, due to reduced input dimensionality. 
Each PINN uses an encoder to project the inputs into a 160-dimensional latent feature space, followed by an output network that predicts the particle concentrations. The activation function is set to ``Tanh''.
A hard constraint is applied to the initial condition of \(c_{-/+}\), which helps improve the model's performance at early time steps~\cite{hassanaly_pinn_2024-1}. The network outputs \(c_{-/+}\) are first scaled to the range $(0,1)$ using a Sigmoid function and then mapped back to the physical concentration domain, following Eqs. \eqref{eq:constrain_ne} and \eqref{eq:constrain_pe}.

\begin{equation}
c_{-}^{\text{enforce}} = \left( \frac{1 - \text{sign}(I)}{2} \cdot (-c_{-}^{\text{init}}) + \frac{1 + \text{sign}(I)}{2} \cdot (c_{-}^{\text{max}} - c_{-}^{\text{init}}) \right) \cdot F(t) \cdot \Hat{c}_- + c_{-}^{\text{init}}
\label{eq:constrain_ne}
\end{equation}
\begin{equation}
c_{+}^{\text{enforce}} = \left( \frac{1 - \text{sign}(I)}{2} \cdot (c_{+}^{\text{max}} - c_{+}^{\text{init}}) + \frac{1 + \text{sign}(I)}{2} \cdot (-c_{+}^{\text{init}}) \right) \cdot F(t) \cdot \Hat{c}_+ + c_{+}^{\text{init}}
\label{eq:constrain_pe}
\end{equation}

In those equations, \(c_{-/+}^{\text{max}}\) and \(c_{-/+}^{\text{init}}\) are the maximum and initial particle concentrations, respectively. The sign function is used to distinguish between charge and discharge profiles, and is defined as:
\begin{equation}
\text{sign}(I) = \begin{cases}  
-1, & I < 0 \quad \text{(discharge)} \\
1, & I > 0 \quad \text{(charge)}.
\end{cases}
\end{equation}
The time-dependent weighting function \(F(t)\) determines the influence of the initial condition and is defined as:
\begin{equation}
F(t) = 1 - \exp\left(-\frac{t}{10 / T_f}\right),
\end{equation}
where \(T_f\) is the total duration of the input profile. Based on trial-and-error tuning, a time constant of 10 seconds was found effective for capturing the effect of the initial condition.

Training is performed using a two-stage optimization: 5,000 epochs with the Adam optimizer (learning rate 0.001), followed by 60,000 epochs of L-BFGS (learning rate 0.01). Double precision is used throughout to ensure numerical stability, which is particularly important for physics-informed learning.

To guide training and improve accuracy, we provide sparse supervised data generated from PyBaMM simulations. Specifically, we sample 10 distinct $(\varepsilon_+, \varepsilon_-)$ combinations within the range $[0.70, 1.00]$. For each pair, the corresponding solution $c^s_i(r, t; \varepsilon_+, \varepsilon_-)$ is used to construct the data loss term in Eq.~\eqref{eq:pinn3}. These sparse data samples help enhance the network's predictions across the parameter space.

The SOH estimation network takes two inputs: (1) a voltage time series segment from the late-charging stage, and (2) key aging-related parameters identified by P-PINNSPM. 
To effectively extract features from these inputs, two encoders are employed: a temporal convolutional network for the voltage time series, and a multilayer perceptron for the static parameters. Both representations are projected into a shared 20-dimensional latent space. 
The network uses ``ReLU'' activations and is trained with Adam for 10,000 epochs (learning rate 0.001). Unlike the PINN, precision is less critical here---both single and double precision yield comparable performance.

\begin{table*}[h]
    \centering
    \renewcommand{\arraystretch}{1.3}
    \caption{Network architecture and key training settings for the two networks. Vol. = Voltage, Param. = Parameters. $3\times50$ indicates a three-layer network with 50 neurons per layer.} \label{tab:net}
    \begin{tabular}{c|c|c|c}
        \hline
        & \textbf{Items} & \textbf{P-PINNSPM} & \textbf{SOH Network} \\
        \hline
        \multirow{4}{*}{Architecture} & Encoder & $3\times 50$ & Vol.: $3\times (64, 32, 20)$; Param.: $3\times 20$ \\
        & Feature Layer & $1\times 160$ & Vol.: $1\times 20$; Param.: $1\times 20$ \\
        & Output Network & $4\times 150$ & $2\times (64, 32)$ \\
        & Activation Function & Tanh & ReLU \\
        \hline
        \multirow{3}{*}{Settings} & Hard Constraint & Yes & - \\
        & Optimization & Adam: 5k; L-BFGS: 60k & Adam: 10k \\
        & Precision & Double  & Single/Double  \\
        \hline
    \end{tabular}
\end{table*}

\subsection{Optimization algorithm for key parameter identification}\label{sec:de}

To identify aging-related parameters, the P-PINNSPM needs an optimization algorithm that minimizes the discrepancy between simulated and observed voltages. We employ the differential evolution algorithm, a robust global optimizer well-suited for complex, non-convex problems \cite{radaideh2023neorl}. 

\begin{equation}
\text{Fitness} = 1000 \cdot \sqrt{ \frac{1}{N} \sum_{i=1}^{N} \left( V_{\text{sim}}(t_i) - V_{\text{exp}}(t_i) \right)^2 } 
+ \lambda \cdot \sqrt{ \frac{1}{N} \sum_{i=1}^{N} \left( t_{\text{sim}}(V_i) - t_{\text{exp}}(V_i) + \Delta t \right)^2 } 
\label{eq:fitness}
\end{equation}

For each charging profile, the algorithm performs 1,000 fitness evaluations to search for the best parameter combination. The objective function, defined in Eq. \eqref{eq:fitness}, evaluates the difference between simulated and observed voltage trajectories,
where \( V_{\text{sim}}(t_i) \) and \( V_{\text{exp}}(t_i) \) represent the simulated and experimental voltages at time \( t_i \), 
\( t_{\text{sim}}(V_i) \) and \( t_{\text{exp}}(V_i) \) denote the simulated and experimental times corresponding to voltage \( V_i \), 
\( \Delta t \) is a time alignment term introduced to match the endpoints of the time-voltage trajectories, 
\( N=100 \) is the number of interpolation points, 
and \( \lambda=50 \) is the weighting factor to balance voltage and time alignment errors. A scaling factor of 1000 is applied to the first term to convert the voltage into millivolts. 

The fitness function is evaluated over tail-end voltage segments of the charging profiles. For RPT cycles with a C-rate of C/3, a 3600-second tail profile is used. For 0.5 C, a 2400-second segment is used. For 1 C, although a 1200-second would be typical, we employ a 1500-second segment instead to account for multi-step charging strategies that reduce current as SOC increases, resulting in an effective C-rate below 1 C. 
Using tail-end charging segments enhances the practicality of the proposed method, as mid-to-high SOC data is more commonly available in real-world applications, such as electric vehicles.

\section{Results}\label{sec:res}
We evaluate the performance of the proposed P-PINNSPM in three stages. First, we demonstrate that P-PINNSPM accurately predicts key battery variables across the entire aging-related parameter space, enabling rapid and reliable identification of internal states.
Next, we assess the impact of including these internal parameters on SOH estimation. Comparisons between models with and without internal information  show that incorporating inferred internal parameters significantly improves accuracy.
Lastly, we highlight the strong extrapolation capabilities of P-PINNSPM, with the model generalizing well to unseen SOH levels, varying charging profiles, and different operating conditions. 

\subsection{Performance of the proposed P-PINNSPM}
Since P-PINNSPM is designed to infer aging-related parameters across the battery's lifetime, its predictive accuray must be high enough to ensure reliable parameter estimation.
Using the training methods described in Section \ref{sec:Network}, we evaluate the model on a RPT charging profile at C/3 rate. 

\begin{figure*}[h] 
    \centering 
    \includegraphics[width=\textwidth]{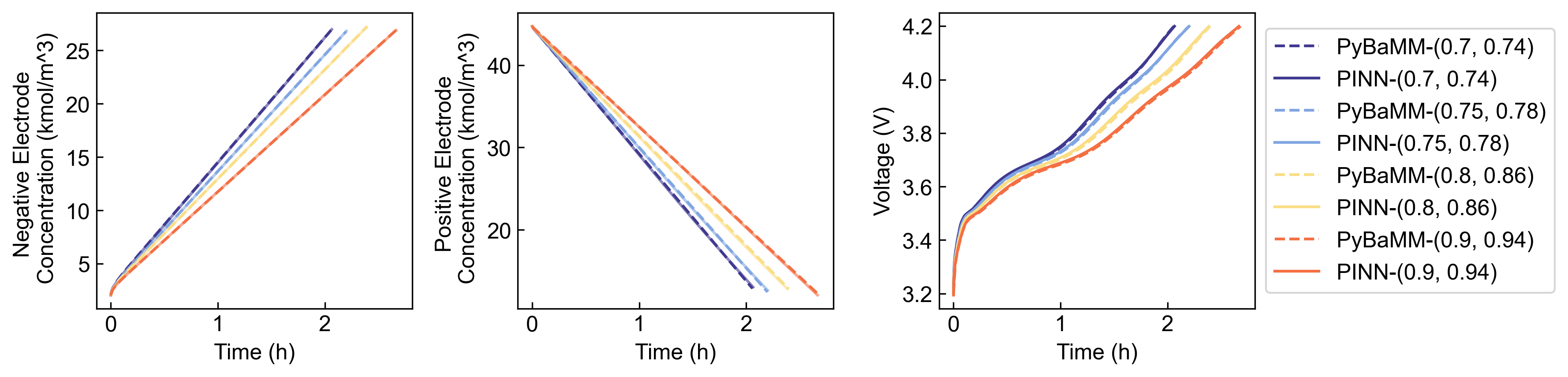} 
    \caption{Prediction performance of P-PINNSPM for concentration and terminal voltage across selected parameter combinations.} 
    \label{fig:iden}
\end{figure*}

The key parameters (i.e., \(\varepsilon_+\) and \(\varepsilon_-\)) are varied within [0.7, 1.0] of their initial values (see Table \ref{tab:params}), which reflects the typical degradation range as cells approach end-of-life.
Figure \ref{fig:iden} shows the predicted concentrations and terminal voltage for four randomly selected parameter combinations. These predictions are benchmarked against results obtained using the finite volume method in PyBaMM \cite{sulzer2021a}.

Across the entire parameter space, the mean relative error is less than 0.3\% for the negative particle concentration and below 0.7\% for the positive particle concentration. The root-mean-square error (RMSE) of the terminal voltage is less than 6 mV. These results show that P-PINNSPM achieves excellent accuracy while generalizing across a wide range of aging-related parameter values. 

\begin{figure}[h] %
    \centering %
    \includegraphics[width=0.56\textwidth]{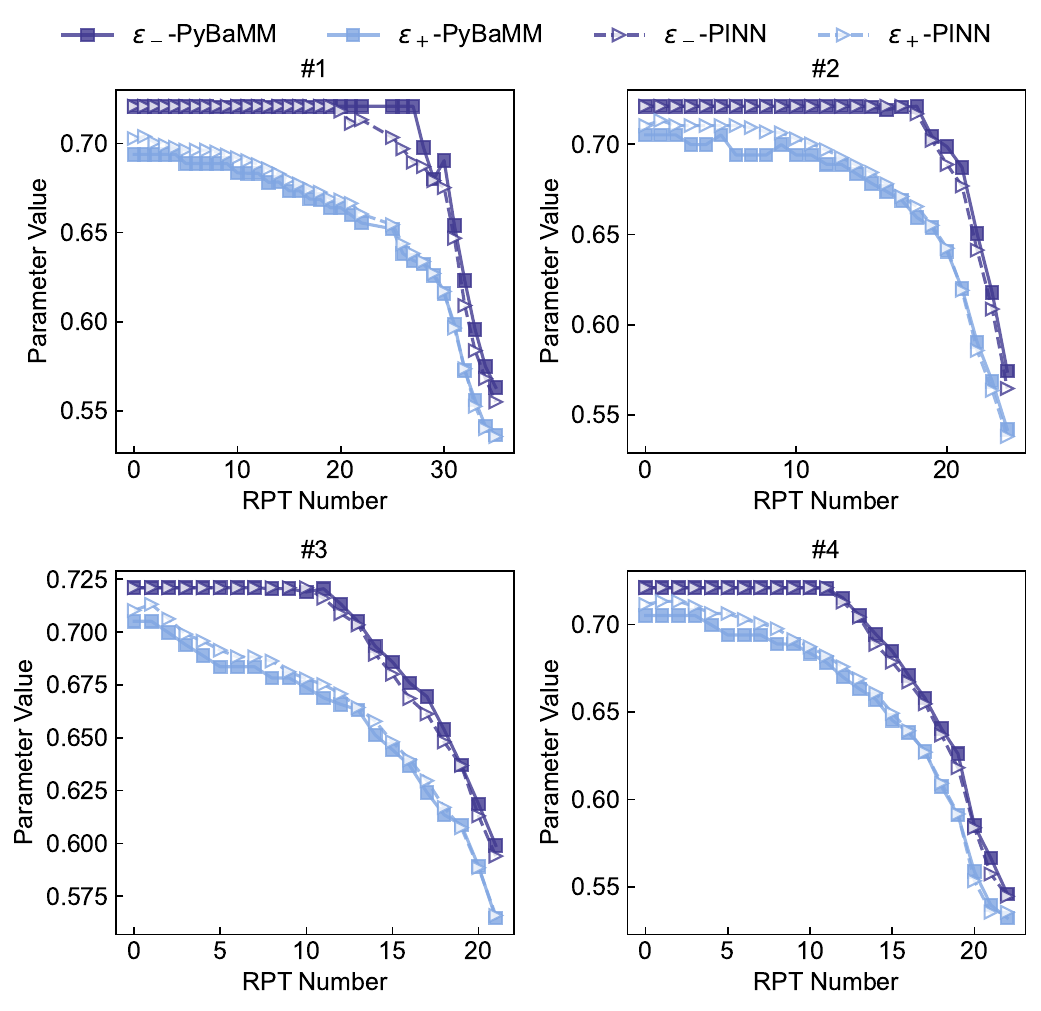} %
    \caption{Comparison of parameter identification results obtained by P-PINNSPM and PyBaMM.} 
    \label{fig:PINN_PBM_IDEN}
\end{figure}

To further evaluate the model's parameter identification capabilities, we compare P-PINNSPM with a conventional SPM model implemented in PyBaMM. 
Both models use differential evolution (as described in Section \ref{sec:de}) to identify parameters based on the final 3,600 seconds of voltage data from  RPT constant current (CC) charging profiles. Figure \ref{fig:PINN_PBM_IDEN} shows the identified parameters for four different cells. Across all cases, P-PINNSPM produces parameter estimates that closely match those from PyBaMM, with mean relative errors below 1\%.

In addition to accuracy, P-PINNSPM offers a significant speed advantage. 
On the same desktop system (12th Gen Intel(R) Core(TM) i7-12700KF CPU),
we ran 1,000 differential-evolution fitness evaluations (10 iterations, each with a population size of 100) for both P-PINNSPM and PyBaMM to ensure a fair comparison. The identification process by PyBAMM takes 1571.40 seconds. 
In contrast, P-PINNSPM completes the task in just 33.58 seconds---achieving a $47\times$ speedup. This efficiency enables real-time identification of internal cell parameters during battery operation.

While these results are based on RPT cycles, the framework is easily extended to other charging profiles. If the input C-rate changes (e.g., 0.5 C or 1 C), a new P-PINNSPM can be trained using the same method. As a result, all cells in Table \ref{tab:2} can be simulated efficiently, and key internal parameters can be reliably extracted across different conditions.

\subsection{SOH estimation with battery internal information}

The strong performance of P-PINNSPM demonstrates its ability to quickly and accurately identify key internal battery parameters, which can be leveraged to improve SOH estimation.
To evaluate this, we identify the parameters of all cells listed in Table \ref{tab:2} over their lifetime, using tail segments from their cyclic charging profiles. 

For the 0.5 C cases, we use a 2,400-second voltage segment from the late CC stage. For 1 C cases, the equivalent duration would be 1,200 seconds. However, a multi-step CC strategy is applied in the 1 C experiments to reduce Li-plating---starting at 1 C and gradually reducing the current as SOC increases. As a result, we use a 1,500-second tail segment for these cases. This selection of tail-end data reflects a practical design choice, as data with such profiles are more commonly available from real-life applications.

To assess the contribution of internal parameters to SOH estimation, we evaluate three cases across the seven cells:
1) Both voltage and parameter information are provided; 2) only voltage is used; and 3) only parameters are used. 
For each test cell, the remaining six are used as the training set.
To minimize the influence of random initialization, we conduct five trials with different random seeds and report the best-performing result for each case, ensuring a fair comparison.

Table \ref{tab:SOHEst} summarizes the results. 
Across all cells, the lowest mean absolute percentage error (MAPE) is consistently achieved when both voltage and parameter information are used.
Notably, the parameter-only case outperforms the voltage-only case in every cell, highlighting the value of inferred internal physical information. The combination of both input types (Vol. + Param.) yields the best results across the board, improve accuracy from 60.61\% to 87.39\%. 

The improvement arises because the internal parameters provide a stable and interpretable signal correlated with capacity degradation, forming a strong foundation for SOH estimation. Although the parameters do not capture all degradation mechanisms, and are subject to model and identification errors, they reliably reflect key aspects such as active material loss. Voltage data complements this by capturing residual effects, further enhancing prediction accuracy.

Of note, cell \#7 exhibits the poorest performance in the voltage-only setting. This is likely due to the absence of comparable operating conditions in the training set, making it an outlier. However, once parameter information is included, the SOH estimation remains accurate, demonstrating the model’s ability to generalize across diverse working conditions. 
More cross-condition estimation results are provided in Section \ref{sec:extra}.

In addition to improving capacity prediction, the inferred parameters offer interpretable insights into electrode-level degradation, providing more than just a scalar SOH value. 
These results highlight the importance of incorporating internal battery information, especially when only limited tail-end voltage data is available---a common constraint in real-world applications.

\begin{table}[h]
    \centering
    \caption{MAPE of SOH estimation using different input information.}
    \label{tab:SOHEst}
    \begin{tabular}{c|ccc|c}
        \toprule
        Cell & Vol. + Param. & Vol. only & Param. only & Improvement \\
        \midrule
        \#1 & 0.380 & 2.995 & 2.050 & 87.32\% \\
        \#2 & 0.643 & 3.342 & 0.774 & 80.77\% \\
        \#3 & 1.043 & 2.647 & 1.306 & 60.61\% \\
        \#4 & 0.854 & 2.363 & 1.264 & 63.85\% \\
        \#5 & 1.137 & 4.495 & 1.864 & 74.71\% \\
        \#6 & 1.104 & 2.745 & 1.219 & 59.79\% \\
        \#7 & 1.624 & 12.888 & 2.902 & 87.39\% \\
        \bottomrule
    \end{tabular}
    \vspace{2mm} \\
    {\small *Note: Best result from 5 random training initializations.}
\end{table}

\subsection{Extrapolation capabilities of P-PINNSPM across unseen scenarios}\label{sec:extra}
\subsubsection{Unseen SOH estimation} A key advantage of P-PINNSPM is its ability to extrapolate to aging states not in the training data. This stems from the embedded physics and the strong correlation between physical parameters and battery degradation. In contrast to purely data-driven methods that solely use data to learn a mapping from macro features (e.g., voltage time series) to capacity and often fail in out-of-distribution settings, our method generalizes well to previously unobserved SOH levels.

To evaluate this capability, we build linear regression models on identified parameters from the tail segments of charging voltage profiles, using two training ranges: 100\%--85\% SOH and 100\%--90\% SOH. The models are then tested on SOH values below these thresholds, respectively.
Data from four cells operated at 0.5 C is used for this demonstration.
As a benchmark, we also train a purely data-driven model using a temporal convolutional encoder with raw voltage data as input, following the same SOH estimation network architecture described in Table \ref{tab:net}.

\begin{figure}
    \centering
    \includegraphics[width=0.85\linewidth]{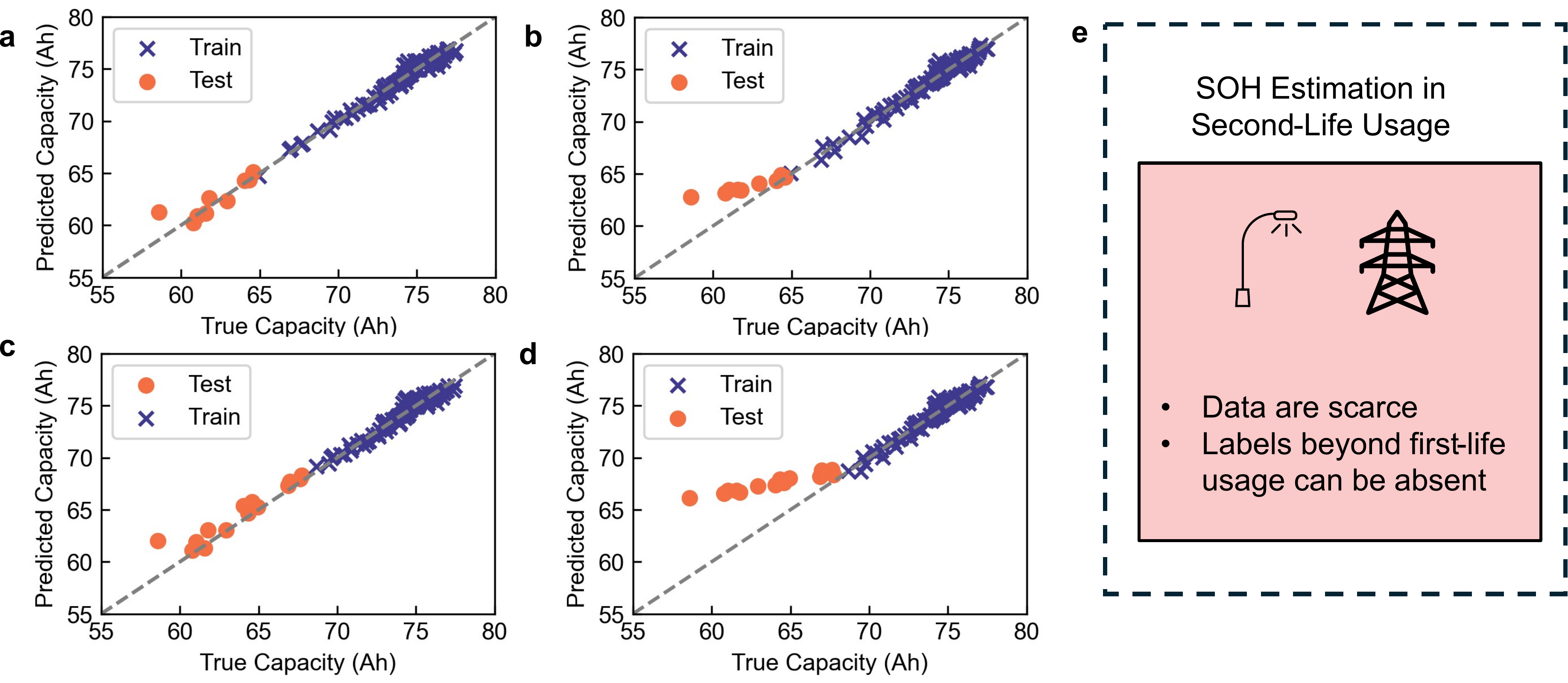} 
    \caption{Extrapolation performance for unseen SOH values. (a): Estimation of SOH below 85\% using identified physical parameters; (b): Estimation of SOH below 85\% using voltage curves; (c): Estimation of SOH below 90\% using physical parameters; (d): Estimatino of SOH below 90\% using voltage curves; (e): Illustration of a potential application scenario enabled by the extrapolation capability, such as second-life battery diagnostics.}
    \label{fig:extrapolateion_sum}
\end{figure}

The results are shown in Figure \ref{fig:extrapolateion_sum}.
When predicting SOH below 85\% (Figures \ref{fig:extrapolateion_sum}(a)--(b)), the linear regression model using identified parameters achieves a MAPE of 1.09\%, while the purely data-driven model using raw voltage input reaches 3.35\%.
When extrapolating below 90\%
(Figures \ref{fig:extrapolateion_sum}(c)--(d)), MAPEs are 1.31\% and 5.91\%, respectively.

\begin{figure}[!t]
    \centering
    \begin{subfigure}[b]{0.36\textwidth}
        \centering
        \includegraphics[width=\textwidth]{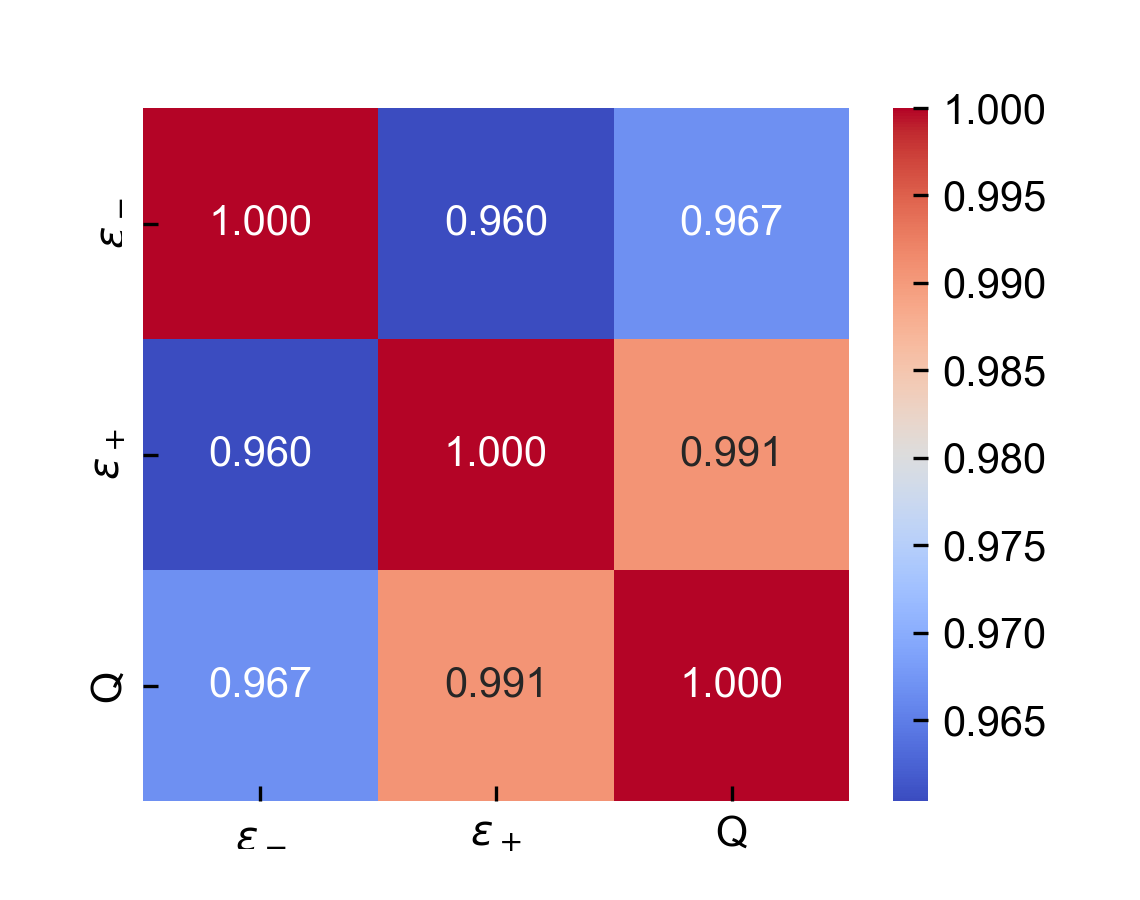}
        \caption{}
        \label{fig:extra_a}
    \end{subfigure}
    \hfill
    \begin{subfigure}[b]{0.42\textwidth}
        \centering
        \includegraphics[width=\textwidth]{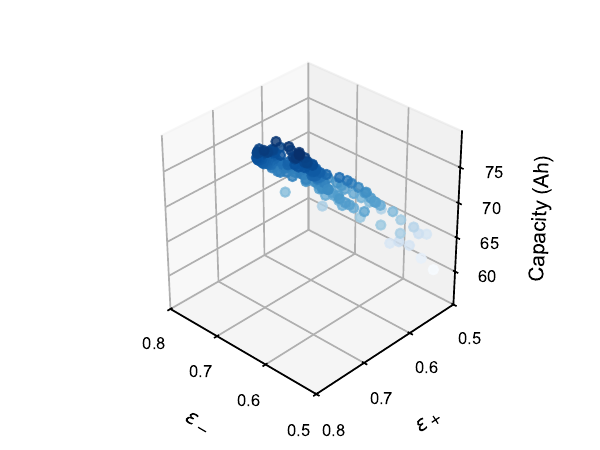}
        \caption{}
        \label{fig:consis}
    \end{subfigure}
    \caption{Parameter–capacity correlation and distribution. (a): Strong correlation between identified parameters and battery capacity; (b): Distribution of key identified parameters across all working conditions, illustrating consistent degradation trends and shared intrinsic parameter space among cells.}
    \label{fig:combined}
\end{figure}

These results highlight the superior extrapolation performance of P-PINNSPM.
The underlying reason is the strong correlation between key aging-related parameters ($\varepsilon_{-/+}$) and capacity degradation. As shown in Figure \ref{fig:extra_a}, both parameters exhibit high Pearson correlation with SOH, effectively capturing the dominating degradation pathways and enabling prediction even beyond training regime.

This extrapolation ability is especially valuable for second-life battery applications (Figure \ref{fig:extrapolateion_sum}(e)), where labeled data is scarce, expensive to obtain, and often unavailable beyond first-life usage \cite{gu_challenges_2024}. By relying on physically meaningful parameters, P-PINNSPM enables robust SOH estimation under such constraints.

\subsubsection{Across-profile generalization within a cell} 
\begin{figure}
    \centering
    \includegraphics[width=0.8\linewidth]{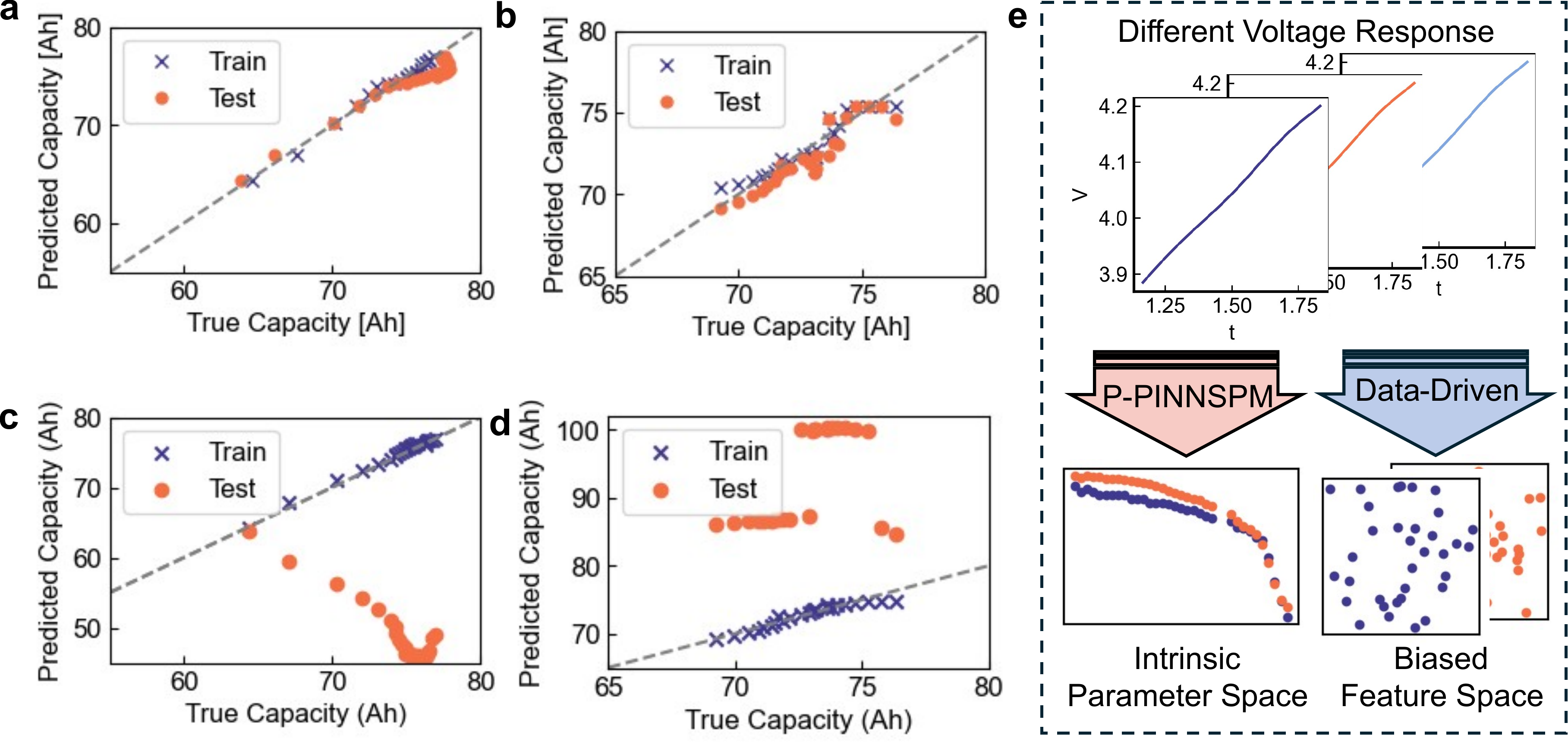} 
    \caption{In-cell SOH estimation performance across charging profiles. (a), (b): Estimation of SOH using inferred parameters for cells \#2 and \#6; (c), (d): Estimation of SOH using voltage curves for cells \#2 and \#6; (e): Illustration of why P-PINNSPM enables reobust cross-profile performance, where voltage responses are mapped to intrinsic, profile-agnostic physical parameters.}
    \label{fig:across_profile}
\end{figure}

We next test whether P-PINNSPM generalizes across different working profiles within the same cell.
To this end, we train the SOH estimation network on RPT charging profiles (C/3) from two cells (\#2 and \#6), and evaluate on tail profiles under 0.5 C and 1 C.
Results are shown in Figure \ref{fig:across_profile}.
The parameter-based approach maintains strong performance across profiles, with MAPEs of 1.580\% and 0.980\%, respectively.
In contrast, the voltage-based method---despite fitting the training data well---exhibits significant errors under new profiles, even for the same cell.

The reason behind these observations is illustrated in Figure \ref{fig:across_profile}(e). While different input profiles produce different voltage responses, the inferred physical parameters remain consistent because they reflect intrinsic material properties. This enables profile-agnostic SOH estimation.
Conversely, the data-driven model maps distinct profiles to different feature spaces, leading to degraded performance in cross-profile settings.

\subsubsection{SOH estimation across working conditions} 
Lastly, we evaluate the robustness of P-PINNSPM across different working conditions. For each condition, we set one as the test set and train on the remaining three.

\begin{figure}
    \centering
    \includegraphics[width=\linewidth]{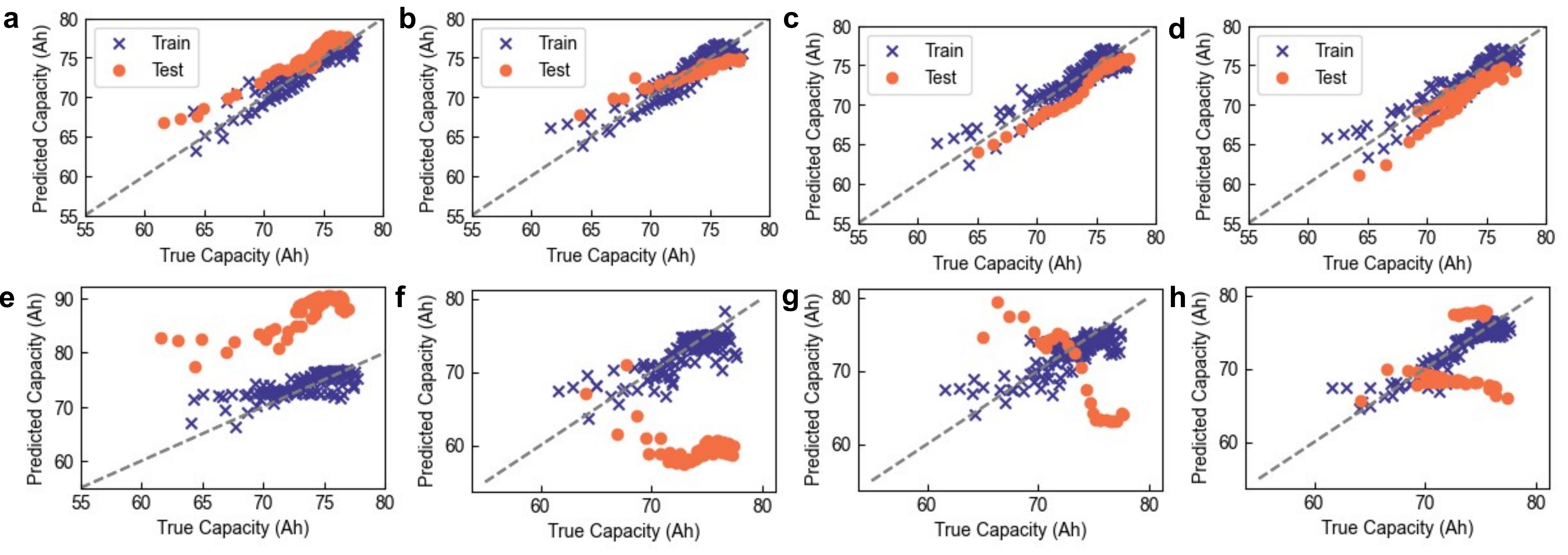}
    \caption{SOH estimation performance across different working conditions. (a)--(d): Extrapolation using identified parameters for various C-rates and SOC ranges: (a) 0.5 C, 0--80\% SOC; (b): 0.5 C, 0--100\% SOC; (c): 1 C, 0--100\% SOC; (d): 1 C, 0--80\% SOC. (e)--(h): Corresponding extrapolation using voltage curves: (e) 0.5 C, 0--80\% SOC; (f): 0.5 C, 0--100\% SOC;  (g): 1 C, 0--100\% SOC;  (h): 1 C, 0--80\% SOC.}
    \label{fig:across_condition}
\end{figure}

Figure \ref{fig:across_condition}(a)--(d) and Table \ref{tab:across_cond} show that the parameter-based model achieves accurate SOH estimates across all conditions, with MAPE ranging from 1.778\% to 2.451\%.
In contrast, the voltage-based data-driven model (Figures (e)–(h)) performs poorly in cross-condition scenarios.
This highlights that the degradation-related parameters evolve consistently under different working conditions, and are projected into a shared intrinsic parameter space (Figure \ref{fig:consis}). As a result, P-PINNSPM generalizes well to new conditions without needing condition-specific training data.
In contrast, voltage-only models suffer from distributional mismatch, as their features are entangled with condition-specific response dynamics.

\begin{table*}[h!]
\centering
\caption{MAPE of SOH estimation across different working condition.}
    \label{tab:across_cond}
    \begin{tabular}{c|cccc}
    \hline
     & 0.5 C, 0--80\% SOC & 0.5 C, 0--100\% & 1 C, 0--80\% SOC & 1 C, 0--100\% SOC \\
    \hline
    P-PINNSPM & 2.451\% & 1.778\% & 2.026\% & 2.375\% \\
    \hline
    Voltage-based & 19.670\% & 18.971\% & 11.791\% & 5.547\% \\
    \hline
    \end{tabular}
    \vspace{2mm} \\
    {\small *Note: Best result from 5 random training initializations.}
\end{table*}

\subsection{Practicality of the propose method}\label{sec:prac}
The proposed method is well-suited for real-world EV applications.
First, Level 2 charging is the most prevalent mode in practice. For example, in the U.S., most home charging uses Level 2, and nearly 80\% of public EV charging ports also fall into this category \cite{afdc_ev_charging}.
Second, Level 2 chargers provide AC charging through 240V (residential) or 208V (commercial) service, typically charging an EV from empty to 80\% in 4--10 hours \cite{usdot_charger_speeds}. This corresponds to a C-rate below 1 C, well within the test range of P-PINNSPM, which has demonstrated reliable SOH estimation under 1 C conditions.
Third, the proposed method uses the tail-end portion of the charging profile, aligning with practical data availability. In real-world settings, EVs are often charged from mid to high SOC, making the required tail-segment data readily accessible. This compatibility with everyday charging behavior enhances the practicability and deployability of the proposed framework.

\section{Conclusions and future work}\label{sec:con}
We developed P-PINNSPM---a parameterized PINN over key aging-related parameters for the SPM. The model accurately predicts battery internal states in approximately 30 seconds, enabling fast and precise identification of critical degradation-related parameters.
These parameters significantly improve battery health estimation, allow extrapolation to unseen SOH levels, and support robust SOH estimation across different charging profiles and operating conditions. These results demonstrate the strong potential of P-PINNSPM for deployment to next-generation battery management systems.

Despite these promising results, some limitations remain. 
First, the current model is based on the SPM and does not account for electrolyte dynamics. As such, it is best suited for scenarios with moderate C-rates (i.e., below 1 C) and is less applicable to high rate or highly dynamic profiles.
Future work will incorporate electrolyte dynamics and variable current profiles to expand applicability to more realistic and demanding scenarios.
Second, the degradation scenarios explored in this study are primarily governed by electrode deterioration, which represents a common degradation pathway. However, batteries can also degrade through other mechanisms such as SEI growth and Li-plating.
To capture these effects, future work should integrate dedicated sub-reaction physics models into the PINN framework.
Including such sub-reactions will further improve the fidelity and generalizability of the model across diverse degradation modes.

\section*{Declaration of Competing Interest}
The authors have no conflict of interest to declare about this work.

\section*{Acknowledgment}
This work is supported by the National University of Singapore Start-Up Grant (A-0009527-01-00) and Farasis Energy USA.

\appendix
\section{Reference Parameter Values for the Single Particle Model}\label{param_val}
\begin{table}[h!]
    \centering
    \footnotesize
    \caption{Reference design and material parameter values for the pouch cells.}
    \begin{tabular}{p{4cm}p{2cm}p{6.5cm}p{1.5cm}p{2.5cm}}
        \toprule
        \textbf{Category} & \textbf{Parameter} & \textbf{Description} & \textbf{Value} & \textbf{Unit} \\
        \midrule
        Geometric parameters 
        & $L_+$ & Cathode thickness & 57.955 & $\mu$m \\
        & $L_\text{sep}$ & Separator thickness & 12 & $\mu$m \\
        & $L_-$ & Anode thickness & 80.12 & $\mu$m \\
        & $A$ & Electrode surface area & 0.02534 & m$^2$ \\
        & $\varepsilon_+$ & Cathode active material volume fraction & 0.714 & - \\
        & $\varepsilon_-$ & Anode active material volume fraction & 0.721 & - \\
        & $R_+$ & Cathode particle radius & 3 & $\mu$m \\
        & $R_-$ & Anode particle radius & 5 & $\mu$m \\
        & $N$ & Number of electrodes in parallel & 78 & - \\ 
        Transport parameters 
        & $D^s_+$ & Cathode solid diffusion coefficient & 4.241e-14 & m$^2$ s$^{-1}$ \\
        & $D^s_-$ & Anode solid diffusion coefficient & 1.135e-14 & m$^2$ s$^{-1}$ \\
        Kinetic parameters 
        & $k_+$ & Cathode reaction rate coefficient & 1.695e-07 & m$^{2.5}$ mol$^{0.5}$ s$^{-1}$ \\
        & $k_-$ & Anode reaction rate coefficient & 2.747e-06 & m$^{2.5}$ mol$^{0.5}$ s$^{-1}$ \\
        Concentration parameters 
        & $c_{s,\text{max}}^+$ & Cathode maximum ionic concentration & 49,034 & mol m$^{-3}$ \\
        & $c_{s,\text{max}}^-$ & Anode maximum ionic concentration & 31,085 & mol m$^{-3}$ \\
        & $c_{s,0}^-$ & Initial cathode concentration & 45,421 & mol m$^{-3}$ \\
        & $c_{s,0}^-$ & Initial anode concentration & 1,554 & mol m$^{-3}$ \\
        & $c_{e,0}$ & Initial electrolyte concentration & 1,000 & mol m$^{-3}$ \\
        \bottomrule
    \end{tabular}
    \vspace{0.5em}
    \begin{minipage}{0.95\textwidth}
    \footnotesize
\vspace{2mm}    \(
    \begin{aligned}
        U^- &= 7.196\!\times\!10^{-13} e^{26.795\,\theta} + 1.124 \!-\! 12.482 \tanh(27.588(\theta\!+\!0.0203)) + 0.334 \tanh(5.231(\theta\!-\!0.3146)) + 1.002 \tanh(32.608(\theta\!-\!0.0998)) \\
        &\quad + 22.562 \tanh(-1.073(\theta\!-\!1.4365)) - 9.089 \tanh(-1.814(\theta\!-\!0.9459)) + 0.930 \tanh(-35.858(\theta\!-\!0.1006))
    \end{aligned}
    \)
    \(
    \begin{aligned}
    U^+ &= -4.407\,\theta + 6.538 + 31.231 \tanh(-4.093(\theta\!-\!0.4354)) + 13.375 \tanh(4.967(\theta\!-\!0.3991)) + 0.564 \tanh(11.457(\theta\!-\!0.2319)) \\
    &\quad - 14.648 \tanh(3.974(\theta\!-\!0.6176)) + 0.478 \tanh(-59.301(\theta\!-\!0.9554)) + 33.344 \tanh(3.636(\theta\!-\!0.5499))
    \end{aligned}
    \)
    \end{minipage}
    \label{tab:params}
\end{table}

\clearpage 
{\scriptsize
\bibliographystyle{elsarticle-num}
\bibliography{cleaned_soh_all}
}

\end{document}